\documentclass[twocolumn,times, tighten]{aastex631}

\journalinfo{Accepted in ApJ | draft version \today}

\usepackage{multirow}

\newcommand{\lya}{Ly$\alpha$}

\newcommand{\ha}{H$\alpha$}
\newcommand{\hb}{H$\beta$}

\newcommand{\ca}{\ion{Ca}{2}}
\newcommand{\cah}{\ion{Ca}{2}~H}
\newcommand{\cak}{\ion{Ca}{2}~K}

\newcommand{\cawav}{\ion{Ca}{2}~8542\,\AA}

\newcommand{\fe}{\ion{Fe}{1}~6302\,\AA}

\newcommand{\RN}[1]{\textup{\lowercase\expandafter{\romannumeral#1}}}

\newcommand{\uat}[2]{\href{http://astrothesaurus.org/uat/#2}{#1 (#2)}}

\usepackage{soul}

\shorttitle{Plage chromosphere with DKIST}
\shortauthors{Kuridze et al.}

\begin{document}

\title{\Large{Insight into the solar plage chromosphere with DKIST}}

\correspondingauthor{D. Kuridze}
\email{dkuridze@nso.edu}
\author[0000-0003-2760-2311]{David Kuridze} 
\affiliation{National Solar Observatory, 3665 Discovery Drive, Boulder, CO 80303, USA}

\author[0000-0002-2554-1351]{Han Uitenbroek} 
\affiliation{National Solar Observatory, 3665 Discovery Drive, Boulder, CO 80303, USA}

\author[0000-0001-6907-9739]{Friedrich W\"oger} 
\affiliation{National Solar Observatory, 3665 Discovery Drive, Boulder, CO 80303, USA}

\author[0000-0002-7725-6296]{Mihalis Mathioudakis}
\affiliation{Astrophysics Research Centre, School of Mathematics and Physics, Queen’s University Belfast, Belfast BT7 1NN, UK}

\author[0000-0002-6547-5838]{Huw Morgan}
\affiliation{Department of Physics, Aberystwyth University, Ceredigion, SY23 3BZ, UK}

\author[0000-0001-5699-2991]{Ryan Campbell}
\affiliation{Astrophysics Research Centre, School of Mathematics and Physics, Queen’s University Belfast, Belfast BT7 1NN, UK}

\author[0000-0001-9352-3027 ]{Catherine Fischer} 
\affiliation{National Solar Observatory, 3665 Discovery Drive, Boulder, CO 80303, USA}

\author[0000-0002-6116-7301]{Gianna Cauzzi} 
\affiliation{National Solar Observatory, 3665 Discovery Drive, Boulder, CO 80303, USA}

\author[0000-0002-7451-9804]{Thomas Schad}
\affiliation{National Solar Observatory, 3665 Discovery Drive, Boulder, CO 80303, USA}

\author[0000-0001-8016-0001]{Kevin Reardon}
\affiliation{National Solar Observatory, 3665 Discovery Drive, Boulder, CO 80303, USA}

\author[0000-0002-3009-295X]{João M. da Silva Santos}
\affiliation{National Solar Observatory, 3665 Discovery Drive, Boulder, CO 80303, USA}

\author[0000-0001-7706-4158]{Christian Beck} 
\affiliation{National Solar Observatory, 3665 Discovery Drive, Boulder, CO 80303, USA}

\author[0000-0003-3147-8026]{Alexandra Tritschler} 
\affiliation{National Solar Observatory, 3665 Discovery Drive, Boulder, CO 80303, USA}

\author[0000-0002-7213-9787]{Thomas Rimmele}
\affiliation{National Solar Observatory, 3665 Discovery Drive, Boulder, CO 80303, USA}


\begin{abstract}
The strongly coupled hydrodynamic, magnetic, and radiation properties of the plasma in the 
solar chromosphere makes it a region of the Sun’s atmosphere that is poorly understood. 
We use data obtained with the high-resolution Visible Broadband Imager (VBI) equipped 
with an \hb\ filter and the Visible Spectro-Polarimeter (ViSP) at the Daniel K. Inouye Solar Telescope 
to investigate the fine-scale structure of the plage chromosphere. To aid the interpretation of the 
VBI imaging data, we also analyze spectra from the CHROMospheric Imaging Spectrometer 
on the Swedish Solar Telescope. The analysis of spectral properties, such 
as enhanced line widths and line depths explains the high contrast 
of the fibrils relative to the background atmosphere demonstrating that \hb\ is an excellent diagnostic for 
the enigmatic fine-scale structure of the chromosphere. A correlation between 
the parameters of the \hb\ line indicates that opacity broadening created by overdense fibrils
could be the main reason for the spectral line broadening observed frequently in chromospheric 
fine-scale structures. Spectropolarimetric inversions of the ViSP 
data in the \cawav\ and Fe {\sc{i}} 6301/6302 {\AA} lines 
are used to construct semiempirical models of the plage atmosphere.
Inversion outputs indicate the existence of dense fibrils 
in the \cawav\ line. The analyses of the ViSP data show 
that the morphological characteristics, such as orientation, inclination and length of 
fibrils are defined by the topology of the magnetic field in the photosphere. 
Chromospheric maps reveal a prominent magnetic canopy in the 
area where fibrils are directed towards the observer.
\end{abstract}



\section{Introduction}
\label{intro}

\begin{figure*}
\centering
\includegraphics[width=\textwidth]{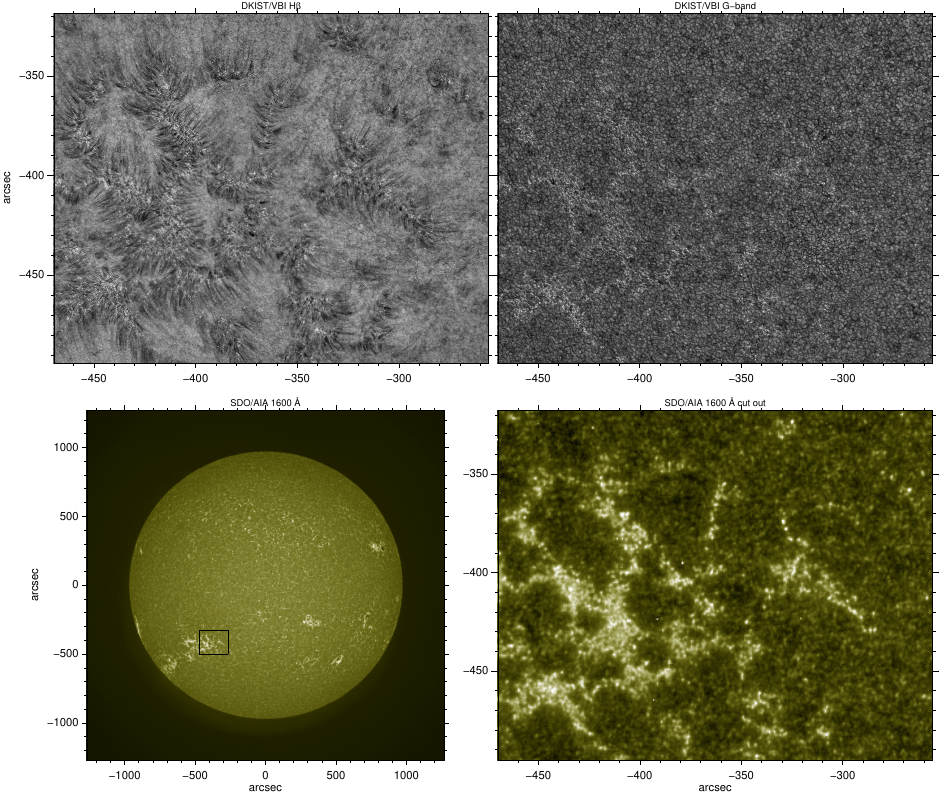}
\caption{Overview of the DKIST observations of the plage.
Top row: Mosaics of high-resolution \hb\ (left panel) and G-band (right panel) filtergrams obtained 
with the VBI instrument on 2022 June 3 between 17:10 and 19:20 UT.
Bottom left: A context  AIA 1600 {\AA} passband full-disk image taken on 2022 June 3 at 16:40 UT. 
The black box indicates the plage region observed with DKIST. 
Bottom right: A cut out of the 1600 {\AA} image coaligned with the DKIST/VBI FoV.}
\label{fig1}
\end{figure*}

The highly inhomogeneous nature of the solar chromosphere was discovered already
in the second half of the 19th century by Father Angelo Secchi, who 
reported the existence of `vertical flames' seen at the solar limb \citep{Secchi1877}, which are now called 
spicules \citep{Roberts1945}. This makes chromospheric fine-scale structures
one of the oldest scientific topics in solar physics. 

Some major breakthroughs in our 
understanding of spicules, and their on-disk counterparts, mottles,
were delivered with the commissioning of half-meter class solar telescopes 
equipped with narrow-band filters and spectrographs able to perform 
simultaneous imaging and spectroscopic observations \citep[see the reviews by][]{Beckers1968,Beckers1972}.
Such data made possible to perform first high-resolution measurements of spicules characteristics such as 
 morphology, thermodynamical properties, spectral characteristics etc.
The seeing-free broadband imaging provided by the Solar Optical Telescope (SOT) onboard Hinode 
and the adaptive optic systems installed on 1-m class solar telescopes equipped with 
Fabry-P\'{e}rot interferometers such as the Swedish Solar Telescope (SST), 
the Dunn Solar Telescope (DST), and the Goode Solar Telescope (GST) have 
further advanced our understanding of the chromosphere at fine scales. 
For example, the data obtained from these telescopes have identified  more energetic, and short-lived 
features such as type II spicules \citep{dePontieu2007} and their on-disk counterparts, 
Rapid Excursions (RE) \citep{Langangen2008, vanderVoort2009, Kuridze2015}.

Still, many aspects of the fine-scale chromospheric structures, including their magnetic 
field strength and topology, thermodynamical properties and formation 
mechanisms remain poorly understood \citep{Sterling2000,Tsiropoula2012}.
The reasons for this are manifold, and include: i) the very dynamic nature of spicular structures, with
spatial and temporal scales close to the resolution 
of present day solar telescopes; ii)  the large number of structures that appear to overlap; 
iii) the inherently low signal to noise (S/N) ratios of polarimetric datasets in the chromosphere \citep{Bueno2022};
iv) the challenges associated with the modeling and interpretation of chromospheric 
spectral lines that rely on solving the complex non-LTE (i.e., departures from
local thermodynamic equilibrium) radiative transfer problem \citep{Rodrigez2017, Beck2019,Carlsson2019,Kuridze2021,Kriginsky2023}.

\begin{figure*}
\centering
\includegraphics[width=18.1 cm]{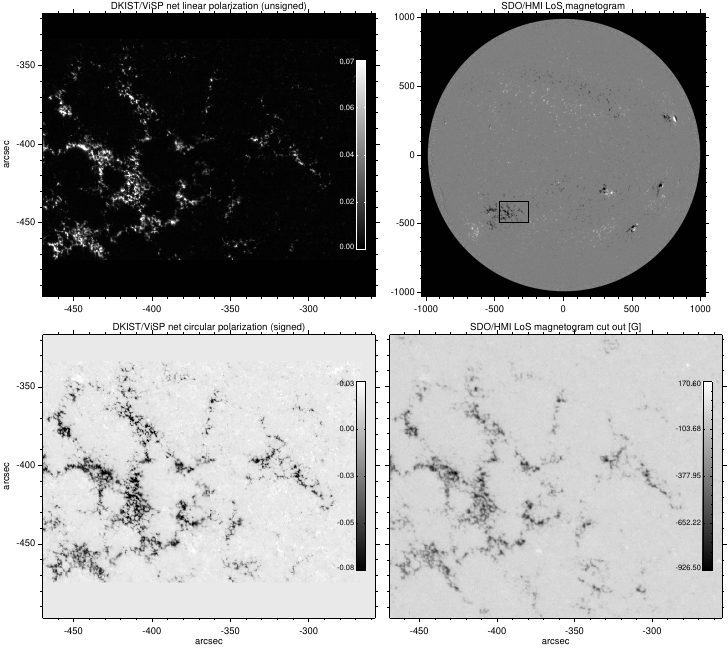}
\caption{Left column: Composite ViSP net linear (top panel) and circular (bottom panel) polarization images compiled from four 
different scans in adjacent positions. The Stokes {\it Q} \& {\it U} and {\it V} parameters 
are integrated over $\Delta\lambda=\pm0.17$~{\AA} from the line core rest wavelength of the \fe\ line.
Top right: A context LoS magnetogram of the full Sun in the Fe {\sc{i}} 6173 {\AA} passband
obtained with HMI. 
The black box indicates the plage region observed by DKIST. 
Bottom right: A cut out of the HMI image coaligned with the DKIST FoV.}
\label{fig2}
\end{figure*}

 \begin{table*}
    \centering
    \caption{Summary of the DKIST instruments and data used in this work. Values refer to a single telescope pointing.}
    \label{tab3}
    \begin{tabular}{ c c c c cc c c  l c D@{\hspace{-2pt}}D c c c c cc c c c}
      \hline
      \hline
      Start time & Instrument &  filter/ & Wavelength & Spatial  & Instantaneous & Time & Sampled & Full map & Number  \\
        UT&               & arm &  range ({\AA})   & sampling  &    FoV   & step  & FoV  &  time & of maps  \\
      \hline  
      \hline
                    17:10:50        & VBI & Blue & \hb\ at 4861.13 & 0.011\arcsec & 45\arcsec$\times$45\arcsec & 3 s & 115\arcsec$\times$115\arcsec & 27 s   & 20  \\ 
                    17:10:22        & VBI & Blue & G-band at 4305 & 0.011\arcsec & 45\arcsec$\times$45\arcsec & 3 s & 115\arcsec$\times$115\arcsec & 27 s   & 20  \\ 
      \hline  
                   \multirow{2}{*}{17:09:55}        & \multirow{2}{*}{ViSP} & {arm 1} &  {6295.2 - 6307.8} & 0.0298\arcsec\tablenotemark{\footnotesize a} & 0.2142\arcsec\tablenotemark{\footnotesize b}$\times$75\arcsec& 3.3 s &107\arcsec$\times$75\arcsec &  27 min & 1  \\ 
                            &  & {arm 3} &  {8535.4 - 8544.9} & 0.0194\arcsec\tablenotemark{\footnotesize a} & 0.2142\arcsec\tablenotemark{\footnotesize b}$\times$48\arcsec& 3.3 s & 107\arcsec$\times$48\arcsec   & 27 min &1  \\ 
 \hline
    \end{tabular}
     \tablenotetext{}{$^a$ pixel scale along slit, $^b$ width of the slit.}
     \label{tab1}
  \end{table*}

Chromospheric fine-scale fibrilar structures
are intricately connected to the local magnetic field distribution and are 
thought to act as tracers for the local magnetic field topology, 
much like apparent coronal loops are used as proxies for the 
magnetic field in the lower corona \citep{Prasad2022}. In the quiet 
Sun (QS) network, chromospheric fine structures create rosettes 
(``bushes''), which are clusters of elongated mottles expanding radially around a 
common center over internetwork regions. Active regions (ARs) are characterized 
by an increased number of magnetic elements
producing a large number of fine scale, elongated features, 
referred to as fibrils when on the disk.
Some of the most interesting fibril-dominated ARs are plages. 
These are regions created by mostly unipolar, almost radial kG fields in the photosphere 
that appear as bright areas, called faculae, in photospheric lines \citep[see, e.g.,][]{Spruit1976, Keller2004, Rezaei2007}.
Resemblance has been reported between QS mottles and plage 
fibrils \citep[see the reviews by][]{Tsiropoula2012}.  
\cite{Foukal1971} investigated the morphological relation 
between chromospheric fine structures observed in regions with 
different magnetic activity. By comparing the observed parameters 
of fine-scale chromospheric structures they concluded that there is a continuous 
morphological progression between QS spicules/mottles and AR/plage fibrils.
\cite{Foukal1971} found that spicules/mottles in QS bushes are less inclined 
with respect to the normal to the surface, while AR/plage fibrils 
are mostly horizontal. The reason for this is that the larger deflection of the 
magnetic field in plage creates a relatively lower-lying horizontal canopy in the 
chromosphere. In a follow up study, \cite{Foukal1971a} concluded that fibrils 
are longer than mottles, with stronger magnetic field strengths, but 
similar temperatures and densities. Similar morphological properties of plage fibrils 
have been reported by \cite{Pietarila2009} who analyzed a high-resolution \cak\ 
filtergram of an AR plage recorded by the SST at disk centre.
\cite{Anan2010} analyzed broadband \cah\ filtergrams of a plage region close to the limb 
obtained by Hinode/SOT and found that spicular jets are shorter in the plage than the QS 
limb spicules, and are characterized by ballistic motions under constant deceleration.
  
The strength and local inclination angle of the magnetic field 
in plage magnetic flux concentrations (MFCs) has long been studied  using 
spectropolarimetric observations. Inversions of photospheric spectral 
lines infer kG field strengths in the plage oriented mostly parallel to the solar normal. 
Inclinations from the local vertical by up to 10$^\circ$-20$^\circ$ have also been reported
\citep{Pillet1997,Topka1992,Bernasconi1995,Buehler2015, Buehler2019}.
By inverting the chromospheric He {\sc{I}} 10830 {\AA} line together with 
the adjacent photospheric Si {\sc{I}} 10827 {\AA} line, \cite{Anan2021} found vertically oriented MFCs 
in the plage photosphere that changed direction, becoming nearly horizontal in the 
chromosphere.  Using \cawav\ observations, \cite{Pietrow2020} inferred the magnetic field strength of a MFC in 
plage chromosphere as 450 G, with an inclination of $\sim$16$^\circ$. 
Using full-Stokes observations in the Mg {\sc{i}} 5173 {\AA}, Na {\sc{i}}  5896 {\AA} and \cawav\ lines, \cite{Morosin2020} 
found that the photospheric magnetic field in the plage region were expanding horizontally toward the chromosphere and forming a volume-filling canopy.
They derived a value of $\sim$658 G for the mean total magnetic field strength of the plage canopy in the chromosphere using the WFA techniques.
From the analysis of some of the chromospheric data used in this paper, and using the Weak Field Approximation, 
\citet{Silva2023} derive an average magnitude of the LOS chromospheric magnetic field in the plage 
of approximately $-$210 G, with clear variations among the fibrils surrounding the plage patches.

One of the main spectral characteristics of chromospheric fine-scale fibrillar
structures in strong chromospheric spectral lines is their broader line width relative to the 
background atmosphere. The causes of this line broadening still constitute a challenge.
The combination of multi-instrument, multi-wavelength optical and UV observations has provided  
evidence that there are transition region (TR) and coronal signatures 
occurring co-spatially and co-temporally with chromospheric fine structures 
\citep{dePontieu2011,Madjarska2011,Henriques2016}, suggesting 
that thermal broadening could be a dominant mechanism of the increased line width in spicular structures. 
However, this remains a controversial subject \cite[cf., e.g.,][]{Beck2016}. 
Non-thermal broadening mechanisms have also been investigated to 
explain the enhanced line width of spicular jets 
\citep{Erdelyiandfedun2007,Jess2009, ZaqarashviliandErdeley2009, Kuridze2012, Kuridze2016}.

In this paper we use data obtained with the high-resolution Visible Broadband 
Imager \citep[VBI;][]{Woger2021} and Visible Spectro-Polarimeter 
\citep[ViSP;][]{deWijn2022} on the recently 
commissioned Daniel K. Inouye Solar Telescope \citep[DKIST;][]{Rimmele2020}
to investigate the fine-scale structure of a plage chromosphere. 
We were able to observe a large field of view and resolve fine-scale structures of 
the chromosphere in broadband \hb\ filtergrams. To aid the 
interpretation of the VBI/\hb\ data, we also analyze \hb\ scans obtained with the  
CHROMospheric Imaging Spectrometer (CHROMIS) on the Swedish Solar Telescope. 
Furthermore, we inverted Stokes profiles obtained with the ViSP instrument in the 
Fe {\sc{i}} 6301/6302 {\AA} and \cawav\ lines with the Departure coefficient aided 
Stokes Inversion based on Response functions \citep[DeSIRe;][]{RuizCobo2022} code
and investigate the connection between the plage magnetic field and the fibrils' characteristics.

\section{Observations and data reduction}
\label{obs}
\subsection{DKIST observational setup}

The observations used in this paper were obtained between 17:09 and 19:20 UT
on 2022 June 3 with the Visible Broadband Imager \citep[VBI;][]{Woger2021}
and Visible Spectropolarimeter \citep[ViSP;][]{deWijn2022}
deployed at the Daniel K. Inouye Solar Telescope \citep[DKIST;][]{Rimmele2020}
under good seeing conditions. The telescope was sequentially pointed at four 
adjacent positions to cover a large plage region located at the south-east part of the solar disk. 
The heliocentric coordinates of the center of the full field-of-view (FoV) covered by the 
observations were  ({\it x, y}) = ($-362$\arcsec, $-406$\arcsec), 
with heliocentric angle $\mu\approx0.843$ (see Figure~\ref{fig1}).
Details about the data and instrumentation are given in the 
next subsections, and in Table~\ref{tab1}.

\begin{figure*}
\centering
\includegraphics[width=\textwidth]{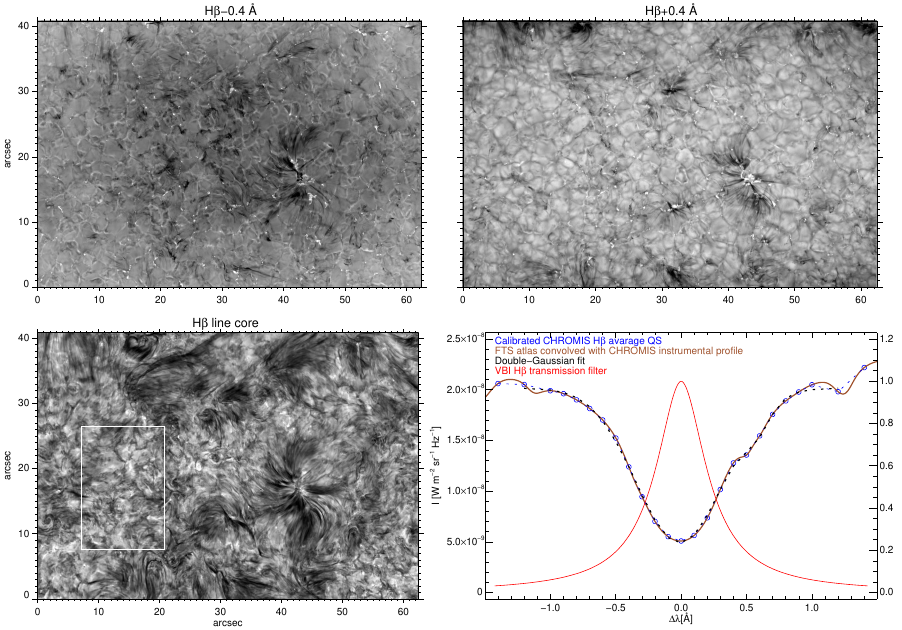}
\caption{Overview of the SST data. Monochromatic \hb\ line 
wing (top row) and core (bottom left) images obtained with the CHROMIS 
instrument at 10:47 UT on 2019 August 3 at disk center.
The bottom right panel displays the experimental transmission profile of the VBI filter (red line, right axis) and the CHROMIS \hb\ average 
line profile (dashed blue line) calculated for a QS region marked with the white box in the bottom left 
panel. Circles show the spectral positions selected for the CHROMIS line scan.
The dashed black line is the double-gaussian fit of the average profile. 
The brown line is the FTS atlas profile convolved with the CHROMIS instrumental profile.}
\label{fig3}
\end{figure*}

\subsubsection{ViSP data}
\label{ViSP}

At each telescope pointing, the ViSP spectropolarimeter performed a large raster 
scan with the 0\farcs2142 wide slit oriented along the solar N-S direction. 
Each scan consisted of 490 slit positions separated by a slit step of 0\farcs219, with 
a time step of 3.3 s, for a total observing time of about 27 minutes per scan. Arms 1 and 3 
of the spectrograph were used to observe the spectral ranges 6295.2 - 6307.8 {\AA} and 8535.4 - 8544.9 {\AA}, 
respectively (arm 2 was not used for these observations). The spatial samplings along the 
slit for arm 1 and arm 3 are 0\farcs0298\,pixel$^{-1}$ and 0\farcs0194\,pixel$^{-1}$,  
with a slit length of $\sim$ 75\arcsec\ and 48\arcsec, respectively.
Hence, the FoV of each raster scan covered 107\arcsec$\times$75\arcsec\ for arm 1 and 
107\arcsec$\times$48\arcsec\ for arm 3  (Table 1). The spectral sampling for 
arms 1 and 3 were 0.0128 and 0.0188 {\AA} per pixel. 

The polarimetric modulation consisted of 10 modulation states that were 
repeatedly observed 12 times each, at a camera frame rate of 41 Hz and with the exposure time of 4 ms for both cameras.
The data were processed 
to level 1 using version 2 of the ViSP data reduction pipeline, that includes polarimetric calibration 
to obtain the full Stokes vector. Additional cross-talk corrections were applied based on the 
method developed by \cite{Jaeggli2022}. More details on the data post-processing can be found in \cite{Silva2023}.


The raster scans at the four different telescope pointings are then stitched together to produce 
large mosaics (Figure~\ref{fig2}) of the observed plage region. Figure~\ref{fig2} shows full maps of \fe\ 
net (wavelenght-integrated) Stokes {\it Q} \& {\it U} and {\it V} coaligned with a cutout of the 
HMI line-of-sight (LoS) magnetogram. Polarization maps indicate that the observed 
region is a highly unipolar plage without a sunspot. 
However, Stokes {\it V} map shows many small-scale parasitic polarity elements that are not resolved in the HMI magnetogram (bottom panels of Figure~\ref{fig2}).

\subsubsection{VBI data}
\label{vbi}

During each ViSP raster scan, we used the VBI blue branch
to obtain filtergrams in G-band, \cak\ and \hb. The wavelengths are selected sequentially 
using interference filters mounted on a rotating filterwheel, and the full width half maximum 
(FWHM) of the filters are approximately 4.37$\pm$0.048, 1.01$\pm$0.02 and 
0.464$\pm$0.01
{\AA} for the G-band, \cak\ and \hb, 
respectively \citep{Woger2014}. Single exposure times were $\sim$0.7, 25 and 4 ms for the G-band, \cak\ and \hb\, 
with a frame rate of up to 30 Hz. In the following, we limit our analysis to G-band and \hb\ data.

The instantaneous FoV of VBI-blue is 45\arcsec$\times$45\arcsec\ with a spatial sampling of 
0\farcs011\,pixel$^{-1}$. For these observations, we employed the so-called ``field-sampling 
mode'' to cover the full DKIST FoV of $2^\prime\times2^\prime$ with a mosaic of 3$\times$3 tiles. 
This fully covers the field sampled by ViSP during the raster {\bf{\citep[cf. Fig. 1 in ][]{Silva2023}}.} All VBI images were reconstructed 
using the speckle code of \cite{Woger2008} to remove residual atmospheric 
distortion from the data, bringing the effective time step to 3 s per tile. Hence, 
at any given wavelength (filter), the full DKIST FoV was sampled in 27 s, and the overall 
cadence for the three filters' sequence was 81 s. For each telescope pointing, 
the sequence was repeated 20 times, for a duration of  $\sim$27 minutes, i.e. the duration of the ViSP raster scan. 
Adjacent tiles in the VBI field-sampling mode have overlaps of about 7\arcsec, and were stitched with sub-pixel accuracy using the World Coordinate System (WCS) header information and cross-correlation techniques. The intensity in overlapping areas is taken as the average at each location.

Finally, these 3$\times$3 mosaics obtained at each of the four different telescope pointings were stitched 
together to produce a large mosaic covering the plage region. 
Adjacent meta-tiles have overlaps of about 20\arcsec\, 
resulting in an overall FoV of about 175\arcsec$\times$205\arcsec\ (Figure~\ref{fig1}). The Figure~\ref{fig1} 
is produced using the tiles with the best spatial resolution.

As mentioned above, seeing conditions were good throughout the observations. After reconstruction, 
we estimate that multiple frames achieved a spatial resolution close to the diffraction 
limit (0.022\arcsec at 430 nm), as determined by the size of the smallest resolved structures seen in the data. 

 \begin{figure*}
 \centering
 \includegraphics[width=\textwidth]{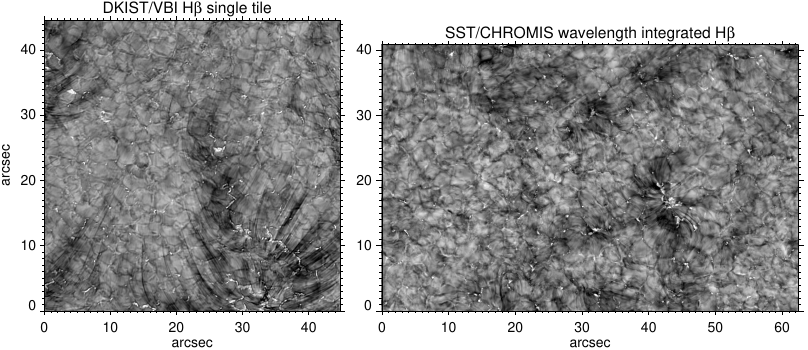}
 \caption{A qualitative comparison of the DKIST VBI image (left panel) and the wavelength integrated 
 SST CHROMIS (right panel) \hb\ image. Before integrating over wavelength, the CHROMIS 
 spectral scan is multiplied by the VBI transmission filter. The DKIST FoV covered by the presented tile is located at around 
 ({\it x, y}) = (-340\arcsec, $-370$\arcsec) in the context image presented 
 in Figure~\ref{fig1} whereas the CHROMIS FoV is located at disk 
 center; they are thus neither co-spatial nor co-temporal}. 
 \label{fig4}
\end{figure*}

\begin{figure*}
\centering
\includegraphics[width=\textwidth]{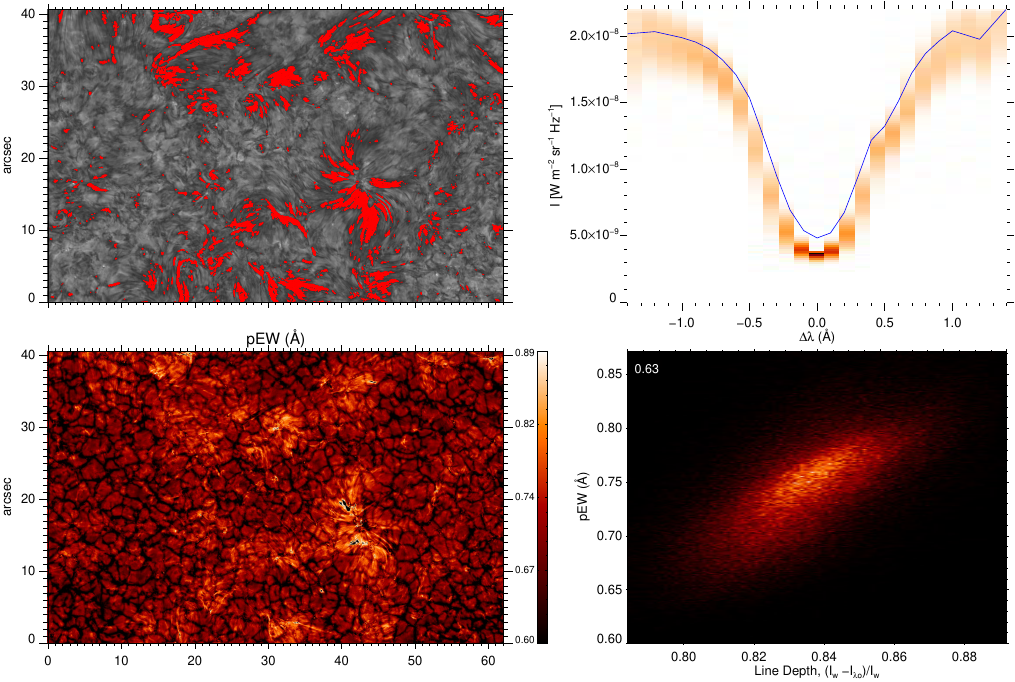}
\includegraphics[width=\textwidth]{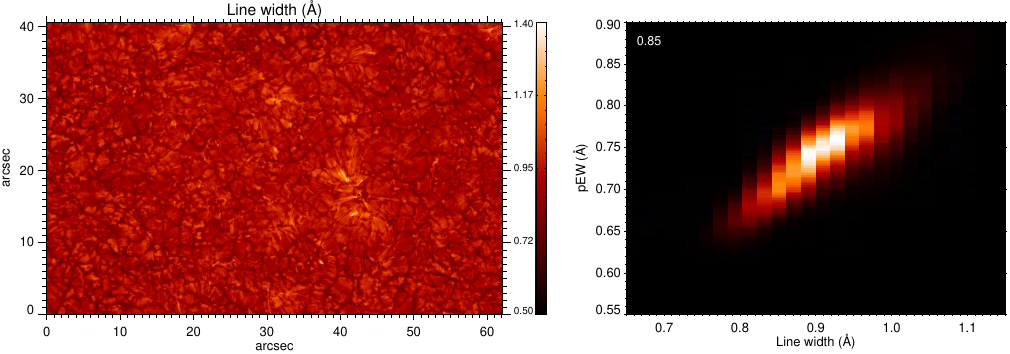}
\caption{Top left: CHROMIS \hb\ line core image overlaid with a mask
where $I^{mean}_{\lambda_0}-I_{\lambda_0}\geqslant 2\sigma$ (see text for explanation). 
Top right: Superimposition of the extracted line profiles of detections shown in the top left panel.
The average \hb\ profile is overplotted as a blue solid line. Middle left: Map of 
pEWs of \hb\ line profiles measured from the double Gaussian model. 
Middle right: Scatter plot of the line core depths vs. pEWs for all identified fibrils.
Bottom left: Map of the width of \hb\ line profiles (see text for details in section~\ref{sp_par}). 
Bottom right: Scatter plot of the line widths vs. pEWs for all identified fibrils. 
The Pearson correlation coefficients are displayed in the top left corners.}
\label{fig5}
\end{figure*}

\subsection{SDO data}

We also analyse 
observations of the same region obtained with NASA’s Solar Dynamics
Observatory (SDO) Atmospheric Imaging Assembly \citep[AIA;][]{Lemen2012} in its 1600 
{\AA} channel and the Helioseismic and Magnetic Imager \citep[HMI;][]{Scherrer2012} 
magnetograms in Fe {\sc{I}} 6173 {\AA}. The SDO observations provide full Sun 
context imaging and allow a comparison with the DKIST data. SDO images are coaligned 
with the DKIST/VBI/ViSP dataset using the SolarSoft {\it auto\_align\_images} 
function via cross-correlation.

\subsection{SST/CHROMIS data}
\label{SST}

Since the VBI data lack fine spectral resolution, we use data obtained with the CHROMIS 
dual Fabry-P\'{e}rot interferometer installed at the SST \citep{Scharmeretal2003a,Scharmer2017} to 
further investigate the spectral characteristics of chromospheric fine-scale structures in \hb. 
Observations were undertaken between 10:47 and 11:04 UT on 2019 August 3 in a quiet area
at disk center. The dataset includes narrow-band spectral imaging in \hb\ with a spatial sampling 
of 0\farcs0378\,pixel$^{-1}$ and a FoV of about 62\arcsec$\times$41\arcsec. Many images in 
the time series were observed under extremely good seeing conditions, with spatial resolution
close to the diffraction limit of the telescope which is $\sim$0\farcs13 at this wavelength. 
The \hb\ line scan consists of 25 profile samples ranging from $-1.2$\,\AA\ to $+1.2$\,\AA\ 
(Figure~\ref{fig3}) from line center. CHROMIS spectral resolution (transmission profile full-width at half-maximum)
is 0.1 {\AA} at 4860 {\AA}. A full spectral scan had an acquisition time of 7 s. The data were processed
using the CHROMISRED reduction pipeline which includes MOMFBD image
restoration \citep{Vannoort2005,Lofdahleta2021}.

The pipeline also applied spectral intensity calibration
to remove the effect of the prefilter transmission profiles
and scale intensity counts to SI units. However, the position angle of the 
science data is not taken into account during the calibration \citep{Lofdahleta2021}. 

For recalibrating CHROMIS \hb\ data we used a spatially averaged
profile over the disk center, quiet-Sun area marked with a white 
rectangle in Figure~\ref{fig3} ($I_{av}(\lambda)$). 
The FTS atlas \hb\ profile ($I_{FTS}(\lambda)$) \citep{Neckel1999} convolved with
the CHROMIS instrumental profile was used as the calibration reference. 
The ratio of the reference and observed intensities 
($I_{FTS}(\lambda)/I_{av}(\lambda)$) was used for calibration and applied to the data (Figure~\ref{fig3}).

\section{Analysis and results}
\label{an_res}

\subsection{DKIST filtergrams}

Figure~\ref{fig1} displays co-aligned images of the plage region in \hb\ and 
G-band obtained between 17:10 - 19:20 UT. Context imaging of the full Sun is 
provided by the AIA 1600 {\AA} filtergram at 17:36 UT. It shows that the 
plage region extends toward the east limb outside the DKIST FoV. The west, 
north and south  borders of the DKIST FoV are surrounded by QS. The bottom 
right panel shows a cutout of the AIA 1600 {\AA} image which is co-aligned 
with the \hb/G-band images.  

The \hb\ image displays multiple features including granulation, small-scale bright points, 
larger scale brightenings and dark fine-scale fibrils. Larger scale brightenings 
coincide with G-band and 1600 {\AA} bright regions suggesting that they 
are associated with strong magnetic flux concentrations (MFCs) known as faculae \citep[\bf{e.g., }][]{Beck2007, Rezaei2007}. 
Faculae are hot granules seen as a result of opacity reduction and 
appear brighter towards the limb \citep{Keller2004,Berger2007}. 
These brightenings are footpoints of plage fibrils appearing as densely-packed, 
and mostly parallel fine-scale strands (Figure~\ref{fig1}). 
The existence of both photospheric and chromospheric features in the \hb\ images is expected
as the VBI/\hb\ filter has a Lorentzian transmission profile centered at the line core 
and covers a significant part of the line wings, with a FWHM of $\sim$0.46 {\AA} 
(bottom right panel of Figure~\ref{fig3}).

\subsection[]{SST/CHROMIS \hb\ images and spectra}

\subsubsection{Overview}

To investigate why and how fine-scale chromospheric features 
appear as high-contrast dark structures relative to the 
background in \hb, we study the line profiles of fibrillar structures using 
high-resolution \hb\ spectral imaging obtained with CHROMIS.
The top and bottom right panels in Figure~\ref{fig3} display CHROMIS \hb\ images in the line 
wings and core obtained at disk center. The FoV is  mostly QS with some network elements and
the rosette region centered around ($x,y$) = (43\arcsec, 15\arcsec) which is dominated by fine-scale 
chromospheric structures. The bottom right panel of Figure~\ref{fig3} shows an average QS \hb\ profile 
calculated over the relatively ``fibril free'' region marked with the white box in the bottom left panel 
of Figure~\ref{fig3} with the spectral positions selected for the line scan.

\subsubsection{Integration with VBI transmission profile}

To more closely replicate the DKIST \hb\ filtergrams, the CHROMIS data were multiplied by the VBI transmission 
profile (bottom right panel of Figure~\ref{fig3}) and integrated over the whole wavelength range 
covered by the spectral scan. Figure~\ref{fig4} shows the comparison of the resulting wavelength-integrated 
CHROMIS and DKIST \hb\ images. For a fair comparison we have selected a 
region of the DKIST FoV covering a significant area of QS. Figure~\ref{fig4} confirms that the wavelength-integrated
CHROMIS \hb\ (called integrated \hb\ from now on)
shows similar scenes including magnetic bright points, granular patterns and 
fine-scale dark fibrillar structures as found in the VBI/\hb\ filtergrams. We note that the granulation in the VBI \hb\ 
image appears more defined. 
This is most likely due to 
the fact that the CHROMIS data do not extend into the  continuum proper, 
but rather only sample up to wavelengths in the wing of the line, formed in the middle photosphere.

\subsubsection{Spectral line parameters}
\label{sp_par}

To study the line profiles of fibrillar structures in the CHROMIS data we 
first identified them in the monochromatic line core images using simple
thresholding. In particular, we calculated the mean line profile, 
$(I_{\lambda_0}^{mean})$ over the ``fibril free''  area marked with the white box in Figure~\ref{fig3}. 
For line core intensities $(I_{\lambda_0})$ we applied a threshold as the mean 
line core intensity ($I^{mean}_{\lambda_0}$) minus 2 standard deviations ($\sigma$).
Resulting detections are presented as red contours plotted over the \hb\ line 
core image in the top left panel of Figure~\ref{fig5}. 

A density diagram of the line profiles 
has been produced with the superimposition of individual profiles from the 
detections (top right panel of Figure~\ref{fig5}). It shows that 
features with increased line depth also show an enhanced spectral width, with respect to the average QS profile
(top right panel of Figure~\ref{fig5}). 

\begin{figure}
\centering
\includegraphics[width=8.7 cm]{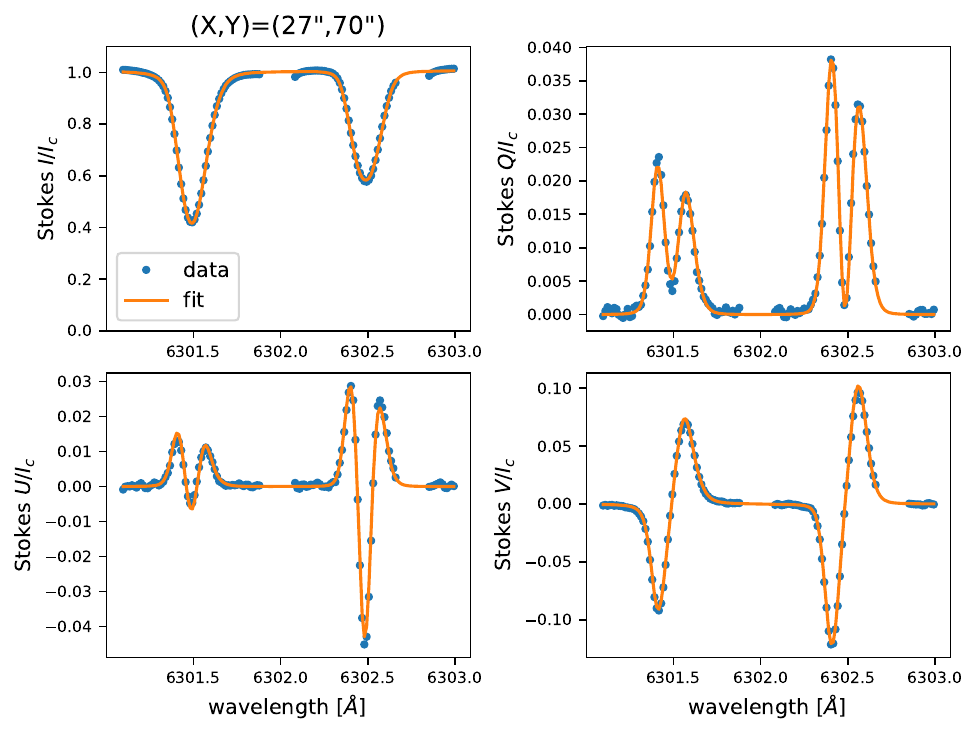}
\caption{A typical set of Fe {\sc{i}} 6301/6302 {\AA} Stokes profiles with corresponding well-fitting synthetic 
profiles obtained from the inversion.
The selected pixel have successfully fitted Stokes profiles with the realistic
atmospheric model. The two O$_2$ telluric lines have been remove from the observed Fe {\sc{i}} 6302 {\AA} spectra.}
\label{fig7}
\end{figure}

We applied a double Gaussian fit to all spectral profiles to take into
account the blend due to the Cr I and Th II near 4861.5 {\AA} (at $\Delta\lambda\approx0.4$ {\AA} 
in the bottom right panel of Figure~\ref{fig3}). We calculated the line depth 
$(I_{\rm{w}}-I_{\lambda_0})/I_{\rm{w}}$ where $I_{\lambda_0}$  
is the intensity of the line core and and $I_{\rm{w}}$ is the
intensity of the line wing located at $\Delta\lambda=-1.2$ {\AA} from the line core.
Due to the absence of a clear continuum reference point in the data, the 
line depth is calculated relative to a position in the far line wing. From the double-Gaussian 
model we also calculated a parameter which we call the pseudo-equivalent width (pEW), 
which is a good proxy for the line equivalent width. The pEW is the wavelength-integrated 
depression of the line profiles for each pixel over the FoV,
\begin{equation}
pEW = \int_{\lambda_{\rm{bw}}}^{\lambda_{\rm{rw}}}\frac{I_{\rm{w}}-I_{\lambda}}{I_{\rm{w}}}\,d\lambda,
\label{eq1}
\end{equation}
where $\lambda_{\rm{rw}}$ and  $\lambda_{\rm{bw}}$  correspond 
to the far red and blue line wing wavelengths located at $\Delta\lambda=\pm1.2$ {\AA} 
from the line core. The middle left panel of 
Figure~\ref{fig5} presents the map of the pEW showing that the detected 
structures have increased pEWs. The middle right panel of Figure~\ref{fig5} 
shows a scatter plot of the pEWs vs line depths for the detected 
fine-structures revealing a correlation between these two parameters 
with a Pearson correlation coefficient of $\sim$0.63. 

\begin{figure*}
\centering
\includegraphics[width=17.5 cm]{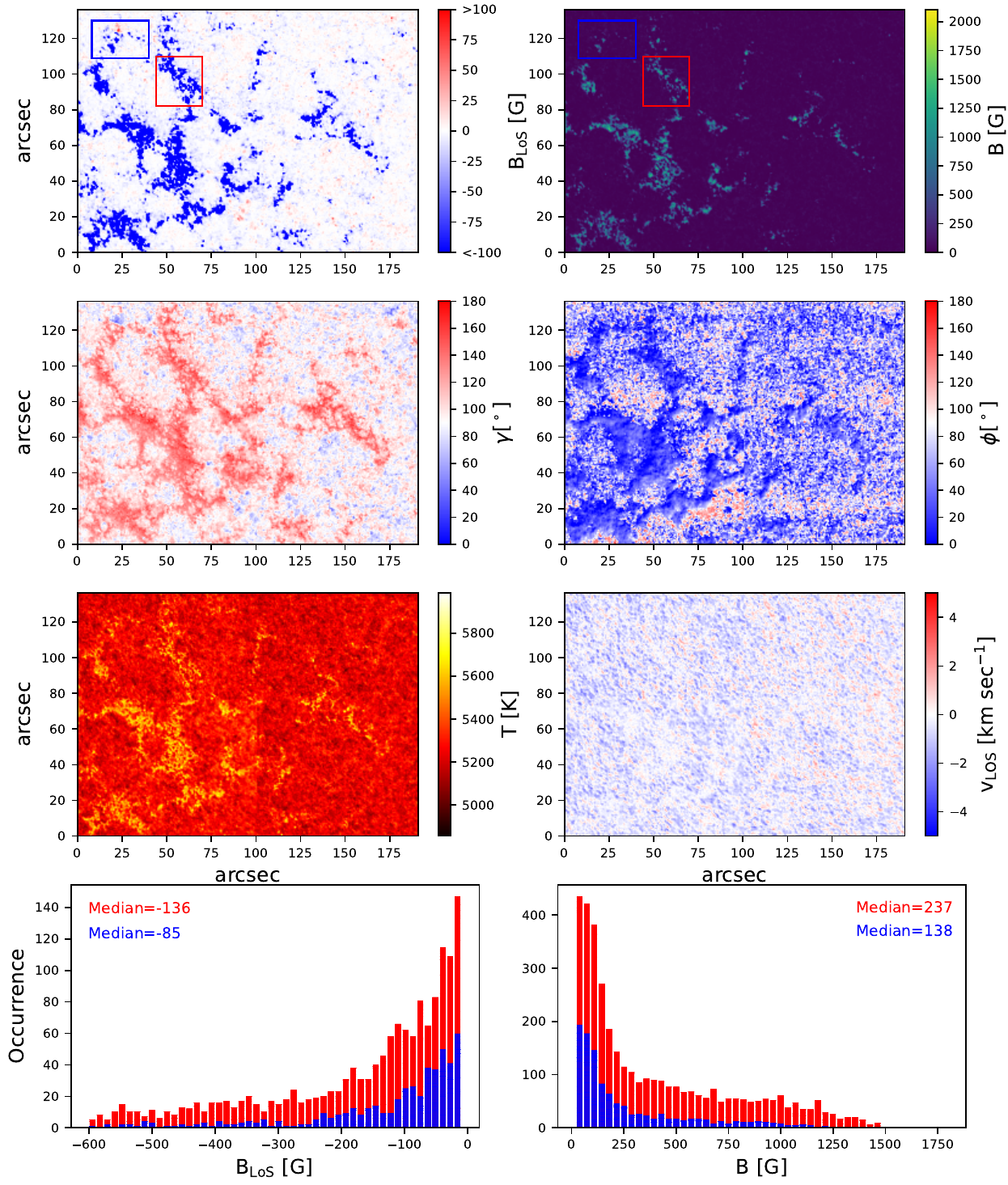}
\caption{Results of the spectral inversions for the Fe {\sc{i}} 6301 and 6302 {\AA} lines, showing the LoS magnetic field, the field strength, 
inclination, azimuth, temperature and LoS velocity maps at $\log\tau$= $-$1. 
The bottom panels depict histograms of the LoS magnetic field and magnetic field 
strength for all pixels with negative $B_{LoS}$} inside the areas marked with the red and blue boxes in the top row. Median values are displayed in the top corners.
\label{fig8}
\end{figure*}

The double-Gaussian models represent well the observed line profiles. However, due to the 
relatively coarse spectral sampling of the data we are not able to resolve and separate Gaussian components 
representing \hb\ and line blends at the red wing, reliably. To quantify the width of the \hb\ line 
profiles from the SST data we followed the method used in \cite{Leenaarts2012}.  For each pixel in the data
we define the width of the line as the intersections of the profile and the line 
$I=I_{min} +(I_{max}-I_{min})/2)$, where $I_{min}$ and $I_{max}$ are profile minimum and maximum. 
The bottom left panel of Figure~\ref{fig5} presents the map of the line width showing 
that the detected structures have increased line width. 
The bottom right panel of Figure~\ref{fig5} shows
scatter plot of the pEWs calculated with the double-Gaussian model 
vs line width. 
There is a clear correlation between 
these parameters suggesting that the pEW and line width are 
increased with the same line broadening mechanism(s) and pEW is a good proxy for the line width.

\subsection{Spectropolarimetric inversions}
\label{inv}
In this section we provide some basic information on the inversions performed on the ViSP 
spectropolarimetric data. A complete description will be presented in a forthcoming paper 
(Uitenbroek et al., in preparation).

\subsubsection{Code setup}

To interpret the ViSP observations, we use the recently developed DeSire inversion 
code \citep{RuizCobo2022}. DeSire uses the LTE inversion code SIR \citep{RuizCoboandiniesta1994}  
with the forward modeling non-LTE radiative transfer solver RH \citep{Uitenbroek2001}.
The code solves the multilevel non-LTE radiative transfer problem 
using analytical response functions derived in LTE to invert or synthesize
photospheric and chromospheric spectral lines in 1D by assuming a plane-parallel geometry.

Prior to the inversion the observed Stokes profiles were averaged 
over 7 and 11 pixels along the slit $(\sim$0.208\arcsec\ and 0.213\arcsec\
for arm 1 and arm 3, respectively) to equalize vertical and horizontal scales of the pixels.
The noise level of the averaged Stokes {\it Q}, {\it U}, \& {\it V} profiles for Fe {\sc} I and \ca\ are 4.4$\times$10$^{-4}I_c$ and 
3.7$\times$10$^{-4}I_{c}$, respectively.
Inversions were performed in four cycles, and Table~\ref{tab2} summarizes the number of nodes used in the cycles.
The magnetic filling factor, or the fraction of a pixel occupied by the magnetic element, is taken as $f=1$.
The stratification of the atmospheric parameters obtained by the inversions 
is given as a function of the logarithm of the 
optical depth scale at 5000 {\AA} (hereafter $\log\tau$).

\subsubsection{Fe {\sc{i}} 6301 and 6302 {\AA} Lines}

We inverted the full Stokes profiles of the Fe {\sc{i}} 6301 and 6302 {\AA} lines over the entire ViSP FoV.
A sample plage profile with signal in all of the polarization parameters is
presented in Figure~\ref{fig7} together with the best-fit synthetic profiles obtained from the inversion. 

\begin{table}
  \centering
\caption{Number of cycles and nodes employed in the inversions of \ca\ and Fe {\sc{i}} lines.}
  \label{tab}
  \begin{tabular}{l   lc}
    \hline
    \hline  
    Physical parameters & nodes used in  \\
       & different cycles \\
    \hline
    %
    %
    Temperature, $T$                  & 2, 5, 7, 9 \\
    Microturbulence, $v_{\rm{mic}}$    &         1, 3, 3, 3 \\ 
    LoS velocity, $V_{\rm LoS}$  &  2, 5, 7, 7 \\
   Magnetic field, $B$    &         2, 5, 7, 7     \\
   Inclination, $\gamma$ & 1, 2, 5, 5 \\
   Azimuth, $\phi$ & 1, 2, 2, 2 \\
    \hline
  \end{tabular}
  \label{tab2}
\end{table}

The inversion output showing maps of the LoS field ($B_{LoS}$), field strength ($B$), 
inclination ($\gamma$), azimuth ($\phi$), temperature ($T$) and LoS velocity 
($v_{LoS}$) at $\log\tau=-1$ (low photosphere) are presented in Figure~\ref{fig8}. 
Azimuth disambiguation has not been performed for the magnetic field, 
and the inclination angle ($\gamma$) and azimuthal angle both describe the direction of the magnetic 
vector with respect to the LoS, with $\gamma=0$ defined as the direction away from the observer.
In the map of $B_{LoS}$ and $B$, we only consider pixels that 
have at least one polarization profile (Stokes {\it Q,~U}  and {\it V})
with maximum absolute amplitude greater  
than 4 standard deviations ($\sigma_s$) \citep{Lagg2016,Campbell2021}.
The inversion reveals a large inclination angle, $\gamma$,
in MFCs. This is due to the large viewing angle (i.e., small $\mu$) of the FoV, where the 
magnetic flux tubes that are oriented mostly along the surface normal are inclined
with respect to the observer. As a result, plage MFCs have a strong magnetic field with 
maximum values of $B$ greater than 2~kG, with a LoS component of 
only a few hundred Gauss  (Figure~\ref{fig8}). Furthermore, the maps reveal 
that the magnetic field of the MFCs inside the plage (see e.g., the area within the red 
box in the top panels of Figure~\ref{fig8}) is stronger than 
the field at the edge of the plage (area within the blue box in the top panels of Figure~\ref{fig8}).

The map of the azimuth shows that there are dominant azimuthal directions 
(dark blue patches in the right panel of the second row of Figure~\ref{fig8}) with 
some variations at the locations of almost all MFCs, confirming the presence of
field lines aligned along the same direction. The azimuth of the magnetic field 
outside MFCs are not well defined as there are very weak {\it Q} and {\it U} signals. 
A temperature map of the inverted region shows temperature
enhancements for MFCs at  $\log\tau=-1$ (Figure~\ref{fig8}).  
The plage elements exhibit enhanced 
temperatures, similar to their appearance in G-band images. 
The LoS velocity map indicates that there is a suppressed granular convection 
(abnormal granulation) in the plage MFCs (Figure~\ref{fig8}).

\subsubsection{Inclination map}

\begin{figure*}
\centering
\includegraphics[width=\textwidth]{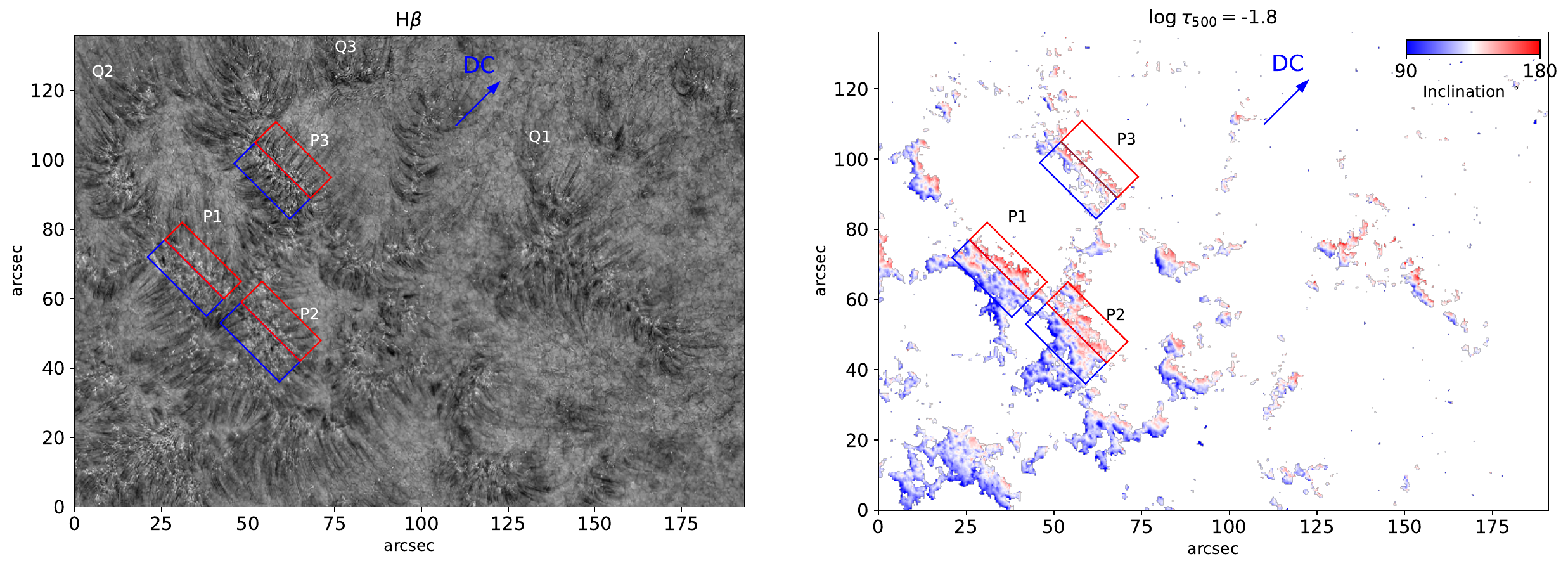}
\includegraphics[width=\textwidth]{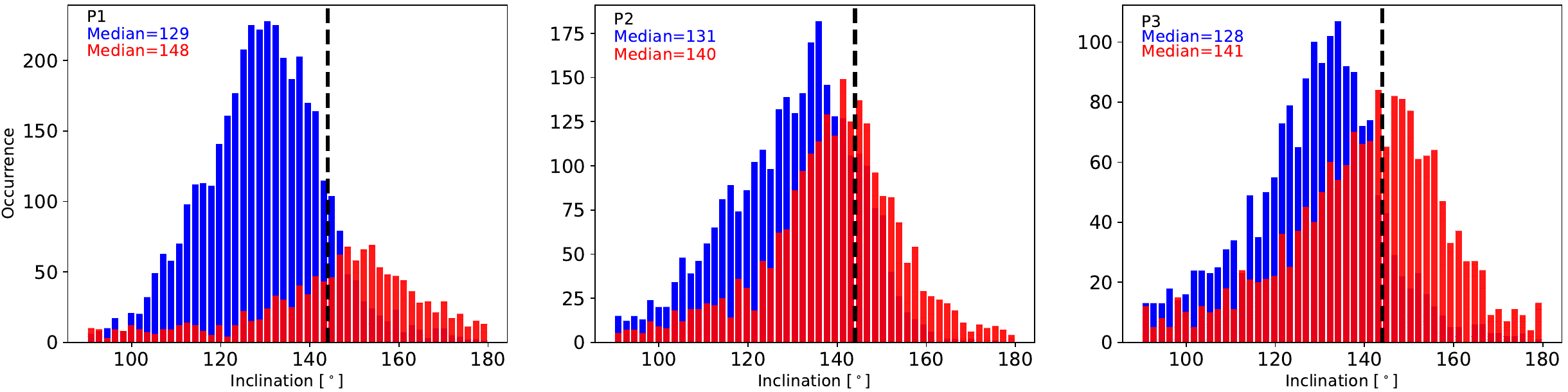}
\caption{Coaligned VBI \hb\ image and inclination map produced with the inversion of 
ViSP Fe {\sc{i}} 6301/6302 {\AA} lines. Q1$-$Q3 mark areas where fibrils are oriented preferably toward the QS.
P1-P3 boxes mark the fibrils that are directed away from the observer (blue boxes) 
and toward the observer (red boxes) inside the plage. Inclination angle histograms of 
the areas marked with blue and red boxes for the P1$-$P3 regions are presented in the bottom 
panels with blue and red bars. Median values are also displayed in the top left corners.
The inclination map and histograms only include pixels 
with a maximum absolute amplitude of the Stokes parameter more than 
$4\sigma_s$ (see text for explanation). The vertical lines mark the inclination angles 
between LoS and local vertical ($\gamma\approx144^\circ$).} 
\label{fig9}
\end{figure*}

Figure~\ref{fig9} shows a co-aligned \hb\ image and 
an inclination map at $\log\tau=-1.8$, 
corresponding to the middle photosphere.
The P1$-$P3 boxes in Figure~\ref{fig9} mark  
selected regions of the MFCs within the plage. The boxes are split into red 
and blue boxes that mark roughly the fibrils directed 
toward the observer (red boxes) and away from the observer (blue boxes).
The disc center (DC) direction is indicated by the blue arrows. 
Histograms of the field inclination within these boxes indicate that magnetic flux tubes associated with the 
lower parts of away-directed fibrils (towards the limb) are less inclined than flux tubes of 
fibrils directed toward the observer (towards the DC; see  bottom panels of Figure~\ref{fig9}).
We'll discuss these findings in Sect. \ref{DC}.

\subsection[]{\cawav\ line}

We invert the \cawav\ (hereafter \ca) together with the photospheric Si {\sc{I}} 8536 {\AA} 
and Fe {\sc{i}} 8538 {\AA} lines located in the blue wing of the \ca\ line.
A five bound level-plus-continuum \ca\ model
atom with complete angle and frequency redistribution is used.
The Stokes {\it Q} and {\it U} signals in the chromosphere 
are inherently weak and the noise level in these data prevent their reliable inversions.

The top row of Figure~\ref{fig10} presents  maps 
of the \ca\ Stokes {\it I} at line
core and Stokes {\it V} at $\Delta\lambda=-0.2$ {\AA}, for the full FoV covered by ViSP scans at four adjacent pointings. 
Due to the shorter length of the ViSP slit in this spectral arm (Table~\ref{tab1})
there is a gap between the upper and lower merged raster scans. 
These maps show that the enhancement of circular polarization signals (Stokes {\it V}) is
cospatial with the MFCs and fibrils directed toward the LoS in the intensity maps 
(Stokes {\it I}) (Figure~\ref{fig10}). 

Representative examples
of Stokes profiles with the best-fit synthetic profiles obtained
from the inversion for the pixels located in a strong MFC and a fibril 
outside MFCs are presented in Figure~\ref{fig11}.
The inversion output showing the maps of temperature, LoS magnetic field,
mass density and LoS velocity, integrated between $\log\tau=-2.8$ and $-4.3$, 
are presented in the middle and bottom rows of Figure~\ref{fig10}. 
The temperature map derived from the inversions (middle row, left column in the Figure)
shows clear temperature enhancements in correspondence of the MFCs, 
but over larger, more diffuse areas than for the photospheric case (Fig. \ref{fig8}).  
The LoS velocity map of the same region reveals weak downflows
along the area where chromospheric fibrillar features are located 
(Figure~\ref{fig10}).

\section{Discussion}
\label{DC}

We investigated the fine scale structure of the plage chromosphere by analyzing 
broadband imaging and full Stokes spectropolarimetric observations 
in the \hb, \cawav, and  Fe {\sc{i}} 6301/6302 {\AA} lines obtained with the 
largest solar optical telescope, DKIST. Plages are unipolar magnetic regions 
appearing either in the vicinity of strong active regions with sunspots or
smaller remnants of a decayed unipolar active regions without sunspots
\citep[\bf{e.g., }][]{Rutten2021}. We observed a non-sunspot-associated plage 
located at around ({\it x, y}) = (-362\arcsec, $-406$\arcsec). 
The large FoV of DKIST covered most of the plage including its boundaries 
with the QS (Figures~\ref{fig1} \& \ref{fig2}).

\subsection[]{The appearance of \hb\ fibrils}

The VBI \hb\ images are dominated by very prominent, fine-scale, 
densely packed fibrils that are mostly parallel to each other (Figure~\ref{fig1})
and are anchored at the unipolar, extended photospheric MFCs (Figure~\ref{fig2}). 
To understand their appearance, we employed 
high-resolution spectral imaging data obtained with SST/CHROMIS on 
a comparable target, and studied their spectral properties (Figure~\ref{fig3}). By multiplying the 
CHROMIS spectral imaging data with the VBI transmission profile and averaging over the 
sampled wavelengths, we are able to reproduce  the \hb\ scene found in the VBI 
data (Figure~\ref{fig4}). The analysis shows that the dark, fibrillar structures seen 
in the composite CHROMIS data have both increased line width and line depth 
with respect to the average QS profile (Figure~\ref{fig5}), explaining why these
features appear as high-contrast dark structures 
relative to the background in the VBI \hb\ image.

\begin{figure*}

\includegraphics[width=8.92 cm]{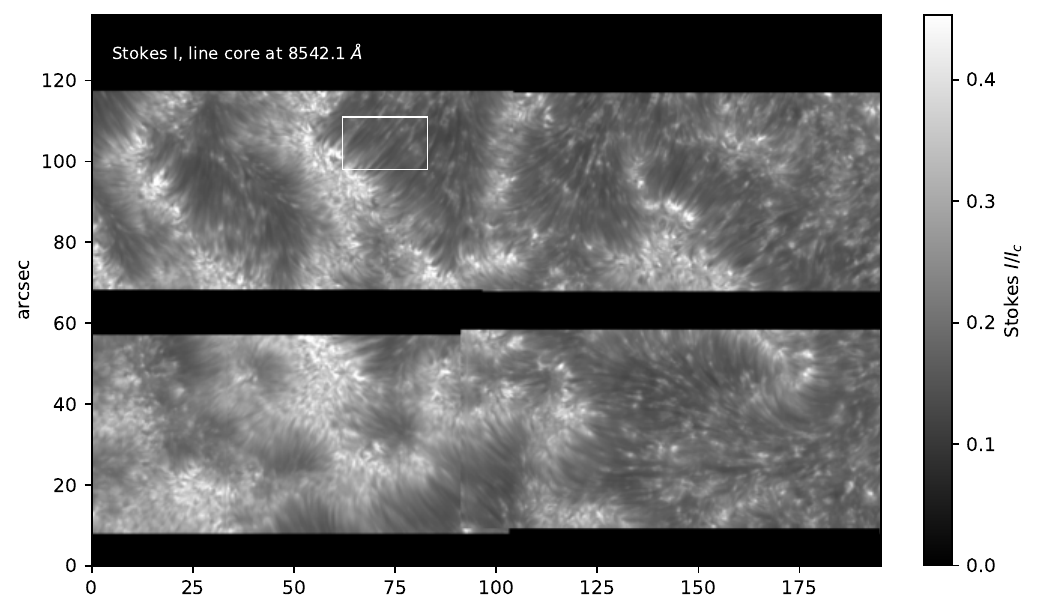}
\includegraphics[width=9.1 cm]{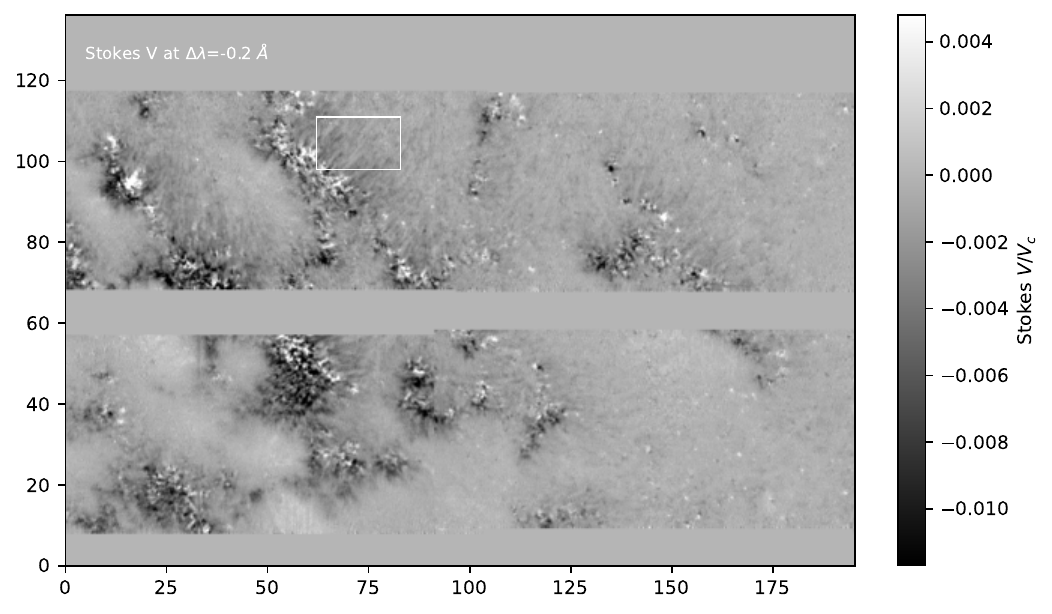}

\includegraphics[width=9.1 cm]{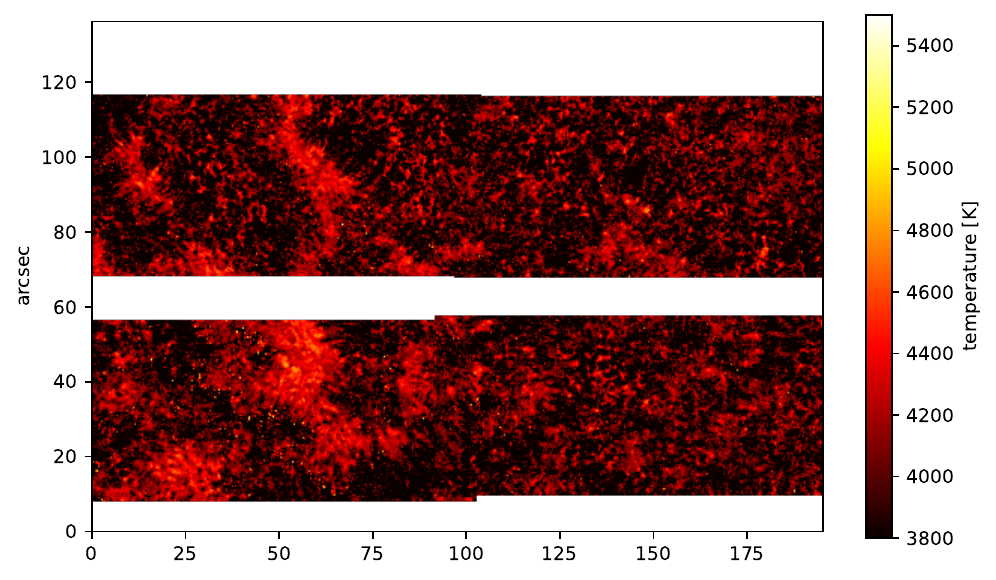}
\includegraphics[width=8.9 cm]{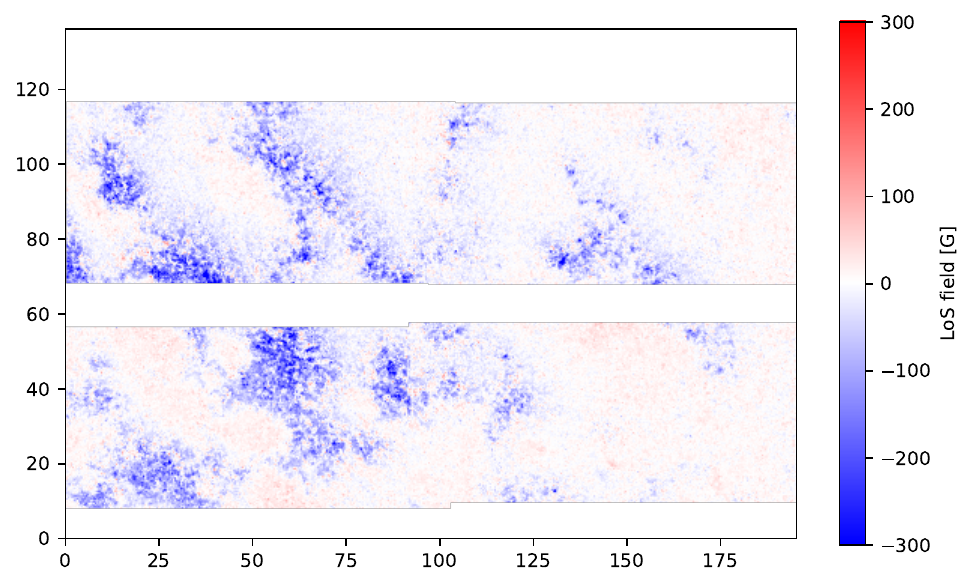}

\includegraphics[width=9.1 cm]{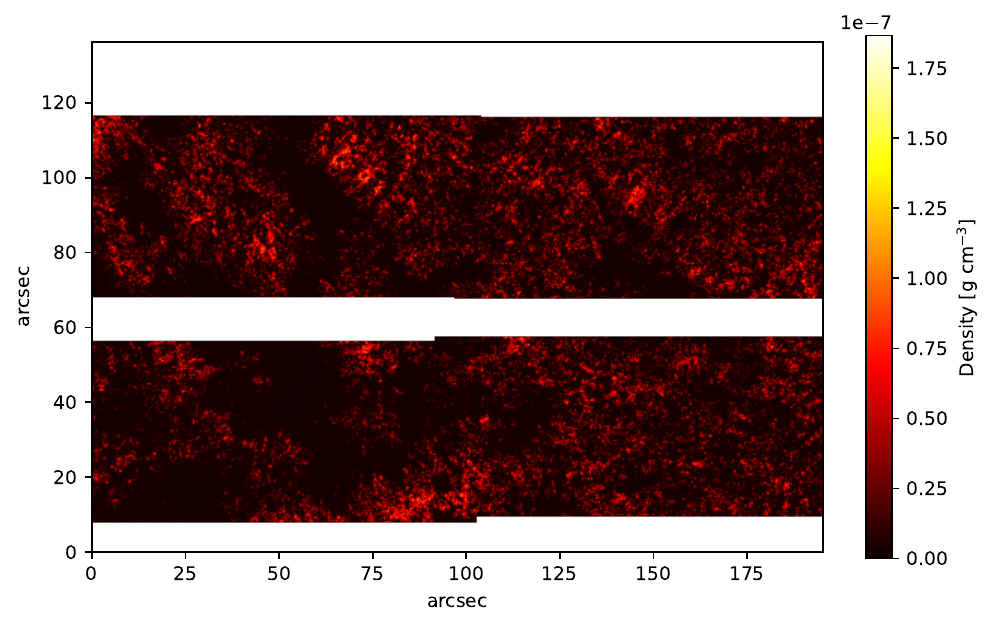}
 \includegraphics[width=8.9 cm]{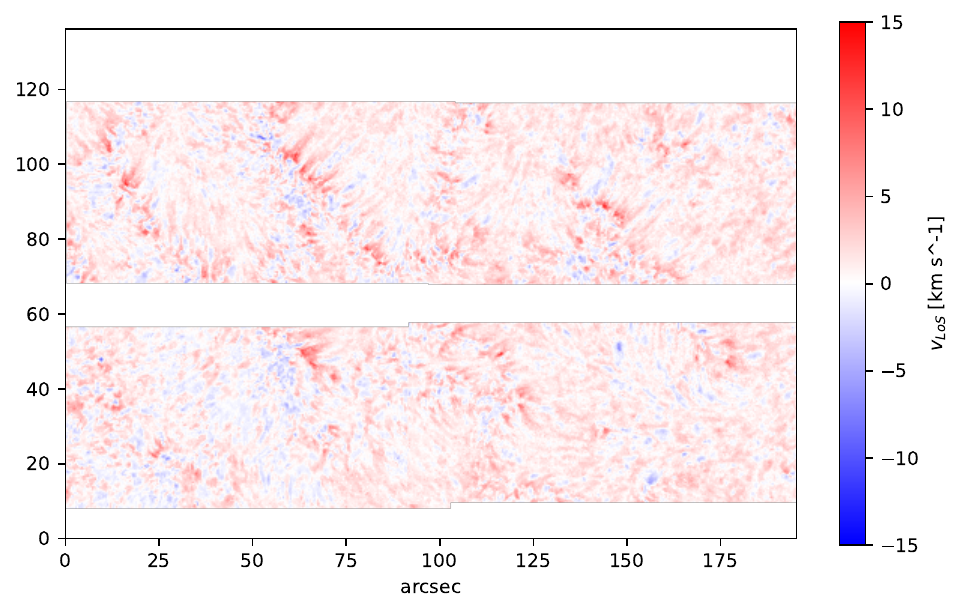}
\caption{The top panels show the ViSP images in the \cawav\ Stokes {\it I} 
line core and Stokes V at $\Delta\lambda=-0.2$ {\AA}. The DeSIRe output showing the temperature, LoS magnetic field, 
mass density and LoS velocity maps averaged over
the interval between $\log\tau=-2.8~ \rm{and} -4.3$ are
presented in the middle and bottom panels.
The white boxes mark an area where fibrils are oriented toward the LoS direction.}
\label{fig10}
\end{figure*}

\begin{figure*}
\includegraphics[width=8.9 cm]{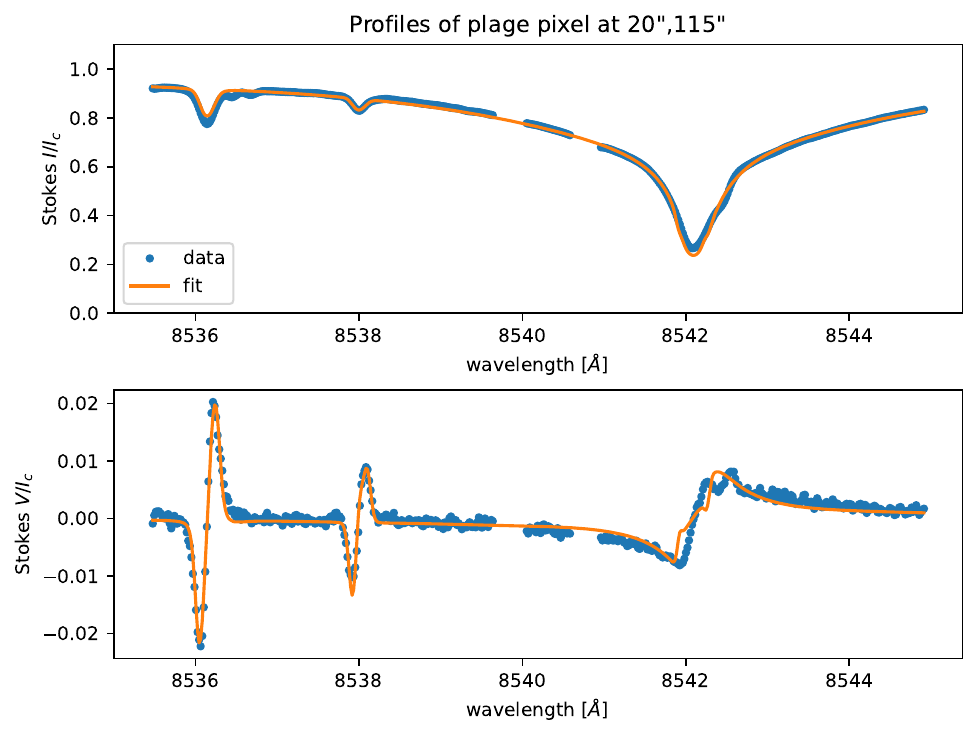}
\includegraphics[width=8.9 cm]{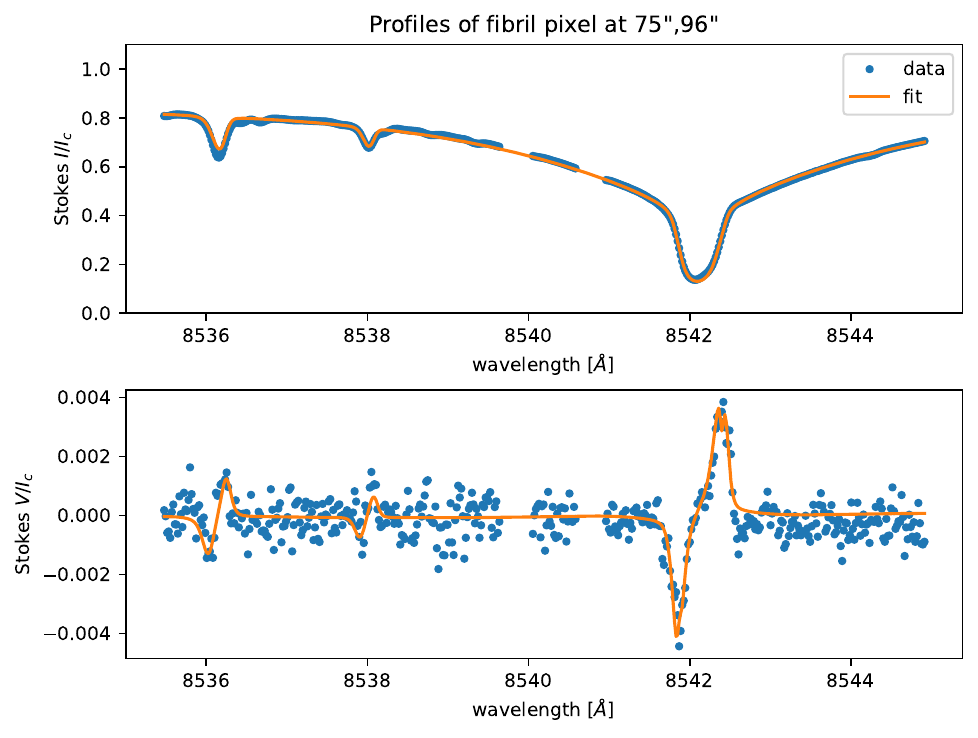}
\caption{Typical set of Stokes profiles with corresponding well-fitting 
synthetic profiles obtained from the inversion of a pixel located in a 
strong MFC (left panels) and a pixel in a fibril outside of MFCs (right panels). These selected pixels have Stokes 
{\it V} successfully fitted with the realistic atmospheric model. The two $\mathrm{H_2O}$ telluric lines have been remove from the observed \cawav\ spectra.}
\label{fig11}
\end{figure*}

\subsection[]{Opacity broadening of the fibrils \hb\ line width}

The line broadening of chromospheric fine structures is a well-known phenomenon 
that has been reported in many observations \citep[see the reviews by][]{Beckers1972,Tsiropoula2012}. 
However, establishing the exact reasons for the line broadening  
remains a central problem of chromospheric research. 
The increased line width can be caused  
by increased local temperatures produced by thermal and nonthermal 
heating processes, nonthermal motions, turbulence  and MHD waves. These mechanisms have been investigated 
in recent decades using both advanced theoretical modeling and high-resolution, 
multi-instrument observations
but a definitive explanation still eludes us.

Another broadening mechanism relevant to chromospheric spectral lines is opacity broadening. 
Increased opacity could enhance the absorption in the line core/inner wing and hence broaden the line width. 
This could also cause larger line core depth if the line core intensity 
is not saturated. The latter scenario is encountered when the observed layer/structure in the chromosphere 
is optically thin \citep{Rutten2003}. In this regime, there is a strong correlation between line 
core optical depth and equivalent width (EW) through the curve-of-growth.

Due to the absence of a clear continuum in our data, 
we are not able to calculate EWs for the \hb\ line. Instead we measured the 
line depression integrated over the wavelength range of the CHROMIS spectral scan. 
This parameter (called pEW) can be taken as a good proxy of the EW and line 
width (Figure~\ref{fig5}). The maps of the pEWs and line width show that the detected fibrillar 
structures have increased pEWs, line widths and variable line core depths (Figure~\ref{fig5}). 
Furthermore, the scatter plot of the pEW vs line depth reveals a clear correlation 
between these two parameters (middle right panel of Figure~\ref{fig5}).
This suggests that the optical thickness of fibrils in the \hb\ line center is less than 1
which classifies the fibril plasma as optically thin in this line, and
not much larger than 1 $(\tau\gtrsim1$). In such conditions, the fibril's line width and 
depth are very sensitive to the optical depth and hence opacity broadening 
can make a significant contribution to the observed line broadening. 
Opacity broadening of plage fibrils in the \ha\ line has been discussed in 
\cite{Molnar2019} where they proposed that the broadening could be 
due to an increased population number of the first excited level of hydrogen, 
due to an enhanced downward \lya\ wing flux.
They concluded that this population number can be affected 
significantly by the downward \lya\ wing flux.

\cite{Leenaarts2012} showed that the \ha\ opacity in the upper chromosphere
calculated using radiation-MHD simulations and three-dimensional non-LTE radiative transfer
computations is mainly sensitive to the mass density and only weakly sensitive to the temperature.
They showed that enhanced chromospheric mass density in fibrils pushes the 
line formation height upward where the source function is set by
the horizontal average of the angle-averaged radiation field. 
On the other hand, the radiation field is independent of the local 
temperature and decreases as a function of height. As a result fibrils appear dark in 
\ha\ against their deeper-formed background \citep{Leenaarts2012}.
Increased opacity in the \hb\ line, manifest as increased line depth and width,
could be related to the increased density in
fibrils, with respect to their background atmosphere. 

In contrast to \hb, no correlation between line core intensity and line 
width of \ha\ has been reported by \cite{Cauzzi2009} (see Figure 6 therein).
This difference is not surprising as the \ha\ is more optically thick. 
The oscillator strength of the \ha\ transition is around 
seven times larger than the oscillator strength of the \hb\ transition.
Therefore, the \ha\ line core intensity for fibrils should 
saturate faster with increased opacity. In our data, we note that despite a clear 
linear tendency, the scattered data points 
(middle right panel of Figure~\ref{fig5}) indicate that the spectra of 
some \hb\ structures can be in the saturation (optically thick) 
regime and/or other broadening mechanisms that are dominant in some fibrils.

\subsection{Morphology of fibrils and Magnetic canopy}

It is widely accepted that fibrils are 
preferentially formed along continuous lines of magnetic force 
with their footpoints  rooted into the photospheric MFCs.
Our observations show that \hb\ fibrils within the plage appear 
shorter compared to fibrils at the edge of the plage, 
parallel to each other, and directed to both sides of the 
extended MFCs (see e.g., regions marked with P1, P2, P3 in Figure~\ref{fig9}).
However, they appear longer and directed preferably toward the 
QS regions near the edge of the plage (see regions marked with Q1, Q2, Q3 
in Figure~\ref{fig9}). Similar results for plage fibrils observed in \ha\ 
and \cak\ filtergrams have been reported  by \cite{Foukal1971, Foukal1971a} and \cite{Pietarila2009}.
Some limitations related to our inversions, such as large viewing angle, 
lack of knowledge of exact geometrical heights, limited FoV, and low S/N for 
linear polarizations in the chromosphere, hinder the determination 
of the 3D geometry of the flux tubes and the height where the magnetic 
canopy is merging in different part of the plage.

\begin{figure*}
\centering
\includegraphics[width=\textwidth]{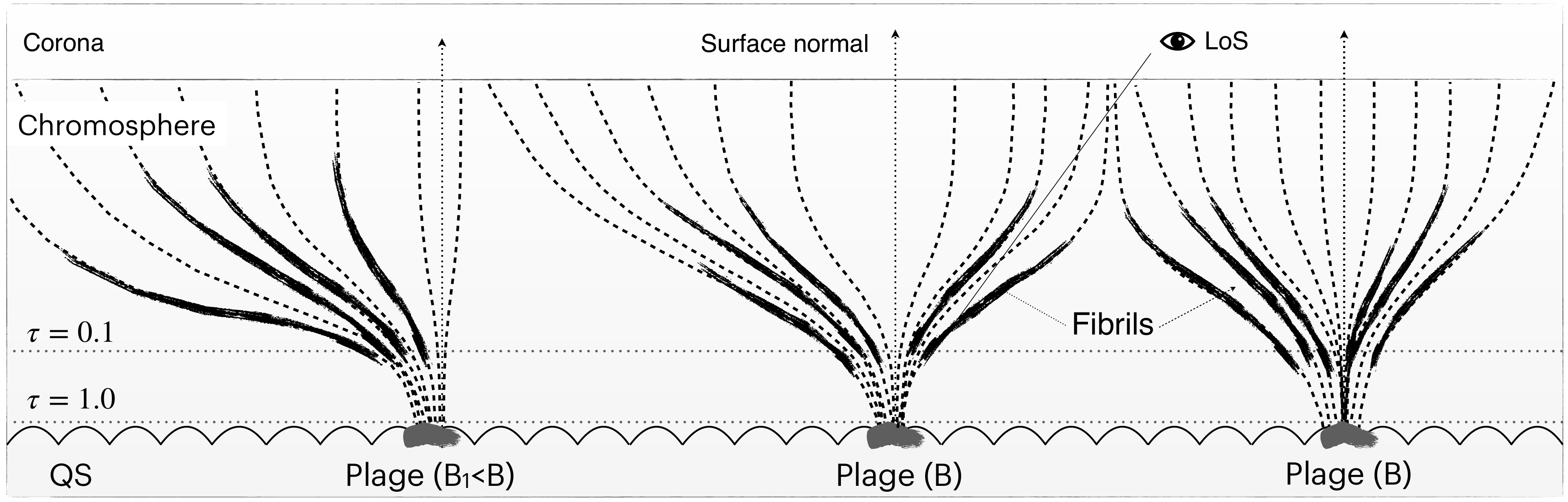}
\caption{Schematic representation of the unipolar plage magnetic field 
configuration showing the basic geometry of the fibrils inside and at the boundaries of the plage. 
Unipolar strong MFCs close to each other inside the plage create low-lying 
canopy/flux tubes with funnel-like cross-sections and shorter fibrils.
At the edges of plage where $B_1<B$, magnetic field lines 
can be pushed away towards the edge of the plage and QS. This creates an asymmetric 
configuration of the funnel-like canopy where flux tubes and fibrils from the edge of the 
plage are directed preferably towards the QS. Dotted horizontal lines mark the 
base of the photosphere ($\tau=1$) and the upper photospheric layer ($\tau=0.1$) 
where Fe {\sc{i}} 6301/6302 {\AA} lines have a high sensitivity.}
\label{fig12}
\end{figure*}

The maps of total magnetic 
field strength and LoS magnetic field in the photosphere at $\log\tau=-1$ 
obtained from the inversion of the Fe {\sc{i}} 6301/6302 {\AA} lines show that there are 
larger MFCs located close to each other inside the plage (Figure~\ref{fig8}), 
hence a smaller separation between MFCs inside the plage than at its edges. 
Following \citet{Durrant1988} and \citet{Pillet1997} we can define 
a ``magnetic filling factor'', $f_l$\footnote{We note the filling factor
introduced here is not the same as the filling factor used in the inversions. 
Here $f_l$ is the large-scale filling factor defined by the separation of MFCs whereas in
the inversions the filling factor is defined within a resolution
element.}, at a reference photospheric height. The canopies of neighboring MFCs merge at a 
distance above the reference level given by the analytical formula, $z_m\approx -2H \ln f_l$, 
where $z_m$  is the height at which the canopy merges,
$H$ is the pressure scale height of the isothermal atmosphere. 
Assuming that the scale height of the plage MFCs is approximately 
the same everywhere, $z_m$ has to be lower inside the plage.
A low-lying canopy with funnel-like cross-section can explain why 
fibrils are shorter inside the plage (Figure~\ref{fig12}).

At the edges of plage regions the magnetic field strength of MFCs are weaker than 
the field strength of neighboring MFCs inside the plage (Figure~\ref{fig8}). 
As a result the funnel-like magnetic field configuration can be bent away from the plage 
toward the edge of the plage and QS to compensate for the lateral magnetic 
pressure imbalance (Figure~\ref{fig12}). This could explain why observed fibrils 
at the edges of the plage have a preferred orientation toward the QS 
(see regions marked with Q1$-$Q3 in Figure~\ref{fig9}).

Maps of the inclination angle reveal that flux tubes at the lower parts of fibrils directed toward the 
observer (toward the DC) have larger inclination angle ($\gamma$) than flux tubes of 
fibrils directed away from the observer (toward the limb) (Figure~\ref{fig9}).
This is manifested in the histograms comparing the distribution of inclination 
angle for blue and red boxes marking lower parts of fibrils directed 
towards the DC and limb, respectively (Figure~\ref{fig9}). 
The average heliocentric angle of the regions marked with P1$-$P3 is  36$^\circ$ ($\mu\approx 0.8$). 
Therefore, flux tubes aligned along the local vertical should have $\gamma\sim144^\circ$ inclination.
The strongest magnetic field patches inside MFCs have a median inclination close to $144^\circ$.
However, the histograms of Figure~\ref{fig9} show that the inclination angle with respect to  $144^\circ$ (local vertical)
for areas marked with red and blue boxes have different distributions and median values. 
This could be an indication that flux 
tubes are beginning to diverge from the vertical at photospheric heights 
as they form a magnetic canopy in the upper atmosphere (Figure~\ref{fig12}).

  \subsection{The chromospheric models}
 
The Stokes {\it V} images of \ca\ at $\Delta\lambda=-0.2$\,\AA\ and the LoS 
magnetic field map obtained from inversion of the \ca\ spectra show that the enhancement in  
circular polarization and $B_{LoS}$ outside the MFCs is detected 
in the areas where fibrils are oriented toward the LoS (North-west) direction 
(see region marked with the white boxes in the top panels of Figure~\ref{fig10}).
Figure~\ref{fig11} compares the Stokes {\it I} and {\it V} profiles of representative pixels 
located on the plage MFC and fibril. The Stokes {\it V} of MFC shows strong 
signal at the Si {\sc{I}} 8536 {\AA} and Fe {\sc{i}} 8538 {\AA} photospheric 
lines as well as in the \ca\ chromospheric line, indicating that the plage 
MFC is dominated by both photospheric and chromospheric magnetic field components.
However, the fibril pixels only show strong Stokes {\it V} for the \ca\ line,
suggesting the presence of purely chromospheric magnetic field. Given the large viewing angle 
for this region, the strong signal in Stokes {\it V} implies a large inclination of the field with respect the local vertical.

Fibrils appear dominated by redshifted plasma in the LoS velocity map (bottom right panel of 
Figure~\ref{fig10}) suggesting the presence of gravity-driven downflows from the top of the fibril 
toward the footpoints. 
The domination of downflows indicates that plage fibrils observed in \ca\ 
are cool and dense plasma which in turn supports the opacity broadening discussed earlier. 
Interestingly, the map of mass density derived from the spectral inversions displays 
areas of enhanced density in the plage fibrils (bottom left panel of Figure~\ref{fig10}). 
Although one has to be cautious in interpreting such a map, as the density is derived from 
the assumption of hydrostatic equilibrium (HE), this represents a tantalizing clue on the nature 
of the fibrils, that will deserve further investigations.

Finally, we note that field aligned, rapid chromospheric plasma downflows  along spicules 
have been recently reported  by \cite{Bose2021}. 
By analyzing their morphological and dynamical properties, \cite{Bose2021} 
suggested that the detected downflows might be the chromospheric counterparts of the frequently observed 
redshifts/downflows predominantly seen in the TR/low coronal lines \citep{Doschek1976,Dadashi2011}.

\section{Concluding remarks}

High-resolution \hb\ filtergrams obtained with DKIST/VBI show
that the plage chromosphere is dominated by fine-scale fibrils. 
Due to their enhanced line width and line depth, the fibrillar features 
are clearly visible in both broadband filtergrams as well as in 
monochromatic line core and wing images. This demonstrates 
that \hb\ observations are an excellent diagnostic to identify 
and track fine-scale structure in the chromosphere.

Our results also suggest that \hb\ fibrils can be optically thin structures,
making this line a very appropriate choice for the well-known cloud modeling 
inversions red proposed originally by \cite{Beckers1964}.
Cloud modeling remains one of the most important inversion techniques 
for the Balmer lines and works at its best for optically thin structures \citep{Tziotziou2007}. 

A correlation between the \hb\ line depth and width indicates that opacity 
broadening could be the main reason for the spectral line broadening observed frequently 
in spicules and fibrils. Whether this mechanism is supported by an increase in electron 
density due to enhanced Ly$\alpha$ flux as proposed by \citet{Molnar2019} or by an 
overall increase in density as proposed by \citet{Leenaarts2012}, remains to be determined. 
It is possible that footpoint heating, manifested as a temperature enhancement of the MFCs 
(Figure~\ref{fig8} and \ref{fig10}) could lead to hot plasma upflows in the chromosphere along magnetic flux tubes;  
rapid cooling of this plasma through radiation losses could then lead to the formation of denser fibrils, 
consistent with the pervasive dowflows observed in the \cawav\ line (Figure~\ref{fig9}). 
Hot plage fibrils footpoints have been reported recently by \cite{Kriginsky2023} using spectral inversions of the same line.
Semiempirical models constructed from the inversions seem to indicate the existence 
of dense fibrils dominated by downflowing plasma in the \cawav\ line.
However, assessing the exact contribution of the opacity broadening in the fibril line widths requires
advanced radiative magnetohydrodynamic computations including 
synthesizing fibrils spectra in \hb. This will be studied in a future paper where we plan to 
perform 3D, non-LTE radiative transfer calculations of the \hb\ line.

The analyses of photospheric and chromospheric polarization data show that morphological 
characteristics, such as orientation, inclination and the length of fibrils, are defined by the topology 
of the magnetic field in the lower solar atmosphere. Future DKIST observations with improved 
S/N for linear polarization signals in chromospheric lines will provide 
vector magnetograms for the 3D geometry of the plage chromosphere.
  
\begin{acknowledgments}
The research reported herein is based in part on data collected with the Daniel K. Inouye Solar Telescope (DKIST) a facility of the National Science Foundation.  DKIST is operated by the National Solar Observatory under a cooperative agreement with the Association of Universities for Research in Astronomy, Inc. 
DKIST is located on land of spiritual and cultural significance to 
Native Hawaiian people. The use of this important site to further scientific knowledge is done so 
with appreciation and respect. DK thanks V. Mart\'{\i}nez Pillet and T. Zaqarashvili for helpful discussions.
DK acknowledge Georgian Shota Rustaveli National Science Foundation project FR-22-7506.
DK acknowledge Science and Technology Facilities Council (STFC) grant ST/W000865/1. MM \& 
RJC acknowledge support from STFC under grants ST/P000304/1, ST/T001437/1, ST/T00021X/1 and ST/X000923/1.
The Swedish 1-m Solar Telescope is operated on the island of La Palma by the Institute for Solar Physics of Stockholm University in the Spanish Observatorio del Roque de los Muchachos of the Instituto de Astrofísica de Canarias. The Institute for Solar Physics is supported by a grant for research infrastructures of national importance from the Swedish Research Council (registration number 2017-00625).

\end{acknowledgments}

\facility{DKIST(VBI \& ViSP), SST(CHROMIS), SDO(AIA \& HMI)}
\software{DeSIRe \citep{RuizCobo2022}, adhoc\_xtalk \citep{Jaeggli2022}}
\bibliography{dkuridze_etal_2023}{}
\bibliographystyle{aasjournal}

\end{document}